\documentclass[useAMS,usenatbib,aas_macros]{mn2e}
\usepackage{aas_macros}

\newcommand{\equ}[1]{eq.~(\ref{eq:#1})}

\newcommand{\Equ}[1]{Eq.~(\ref{eq:#1})}

\newcommand{\se}[1]{\S\ref{sec:#1}}
\newcommand{\fig}[1]{Fig.~\ref{fig:#1}}
\newcommand{\figs}[1]{Figs.~\ref{fig:#1}}
\newcommand{\Fig}[1]{Figure~\ref{fig:#1}}
\newcommand{\Figs}[1]{Figures~\ref{fig:#1}}
\newcommand{\be}{\begin{equation}}
\newcommand{\ee}{\end{equation}}
\newcommand{\bea}{\begin{eqnarray}}
\newcommand{\eea}{\end{eqnarray}}

\newcommand{\no}{\noindent}

\newcommand{\msun}{M_\odot}

\newcommand{\ifm}[1]{\relax\ifmmode#1\else$\mathsurround=0pt #1$\fi}
\newcommand{\kms}{\ifmmode\,{\rm km}\,{\rm s}^{-1}\else km$\,$s$^{-1}$\fi}
\newcommand{\hmpc}{\,\ifm{h^{-1}}{\rm Mpc}}

\newcommand{\sy}{\,\ifm{M_\odot}{\rm yr}^{-1}}

\newcommand{\kpc}{\,{\rm kpc}}
\newcommand{\pc}{\,{\rm pc}}
\newcommand{\Gyr}{\,{\rm Gyr}}
\newcommand{\Myr}{\,{\rm Myr}}
\newcommand{\yr}{\,{\rm yr}}
\newcommand{\ltsima}{$\; \buildrel < \over \sim \;$}
\newcommand{\lsim}{\lower.5ex\hbox{\ltsima}}
\newcommand{\gtsima}{$\; \buildrel > \over \sim \;$}
\newcommand{\gsim}{\lower.5ex\hbox{\gtsima}}

\newcommand{\dd}{{\rm d}}

\def\omm{\Omega_{\rm m}}
\def\oml{\Omega_{\Lambda}}

\def\Ms{M_{\rm d}}
\def\Msd{{\dot{M}}_{\rm d}}
\def\Mv{M}
\def\Rv{R}
\def\Vv{V}
\def\Tv{T_{\rm vir}}
\def\Mvd{{\dot{M}}}

\def\Mhd{\dot{M}_{\rm hot}}
\def\Mh{M_{\rm hot}}
\def\Mg{M_{\rm gas}}
\def\Mps{M_*}
\def\mps{M_*}
\def\Msh{M_{\rm sh}}

\def\tv{t_{\rm vir}}
\def\zb{z_{\rm b}}
\def\fb{f_{\rm b}}
\def\Bv{B_{\rm v}}
\def\Bb{B_{\rm b}}

\def\tq{t_{\rm q}}
\def\tb{t_{\rm b}}
\def\zb{z_{\rm b}}
\def\zq{z_{\rm q}}

\def\dW{\Delta\omega}

\title[Bursting and Quenching in Massive Galaxies]
{Bursting and Quenching in Massive Galaxies\\ without Major Mergers or AGNs}

\author[Y. Birnboim, A. Dekel \& E. Neistein]
{Yuval Birnboim, Avishai Dekel and Eyal Neistein\\
\\
Racah Institute of Physics, The Hebrew University, Jerusalem 91904 Israel\\
yuval@phys.huji.ac.il; dekel@phys.huji.ac.il; eyal\_n@phys.huji.ac.il}

\begin{document}

\large

\pagerange{\pageref{firstpage}--\pageref{lastpage}} \pubyear{2002}

\maketitle

\label{firstpage}

\begin{abstract}
We simulate the buildup of galaxies by spherical gas accretion through 
dark-matter haloes, subject to the development of virial shocks.  We find that 
a uniform cosmological accretion rate turns into a rapidly varying disc buildup
rate.  The generic sequence of events (Shocked-Accretion Massive Burst \& 
Shutdown: SAMBA) consists of four distinct phases: (a) continuous cold 
accretion while the halo is below a threshold mass $\Msh\!\sim\!10^{12}\msun$,
(b) tentative quenching of gas supply for $\sim\!2\Gyr$, starting abruptly 
once the halo is $\sim\!\Msh$ and growing a rapidly expanding shock, 
(c) a massive burst due to the big crunch of $\sim\!10^{11}\msun$ gas in 
$\sim\!0.5\Gyr$, when the accumulated heated gas cools and joins new infalling 
gas, and (d) a long-term shutdown, enhanced by a temporary shock instability
in late SAMBAs, those that quench at $z\!\sim\!2$, burst at $z\!\sim\!1$ 
and end up quenched in $10^{12-13}\msun$ haloes today. 
The quenching and bursting occur at all redshifts in galaxies of baryonic mass 
$\sim\!10^{11}\msun$ and involve a substantial fraction of this mass. 
They arise 
from rather smooth accretion, or minor mergers, which, unlike major mergers, 
may leave the disc intact while being built in a rapid pace. 
The early bursts match observed maximum starbursting discs at $z \gsim 2$,
predicted to reside in $\lsim10^{13}\msun$ haloes. The late bursts resemble 
discy LIRGs at $z \lsim 1$. On the other hand, the tentative quenching gives 
rise to a substantial population of $\sim\!10^{11}\msun$ galaxies with a 
strongly suppressed star-formation rate at $z\!\sim\!1$-$3$.  
The predicted long-term shutdown leads to red \& dead galaxies in groups.  
A complete shutdown in more massive clusters requires an additional quenching 
mechanism, as may be provided by clumpy accretion.  Alternatively, the SAMBA 
bursts may trigger the AGN activity that couples to the hot gas above $\Msh$ 
and helps the required quenching.  
The SAMBA phenomenon is yet to be investigated using cosmological simulations.
\end{abstract}

\begin{keywords}
{shock waves ---
accretion ---
galaxies: evolution ---
galaxies: formation ---
galaxies: haloes ---
dark matter}
\end{keywords}

\section{Introduction}
\label{sec:intro}

Observations reveal a puzzling phenomenon of ``maximum starbursting" in
massive galaxies at high redshift in cases where major mergers
are ruled out.  Examples include LIRGs at $z \lsim 1$ \citep{hammer05}
and maximum bursters at $z \gsim 2$ \citep{genzel06,forster06}.
In these cases,
galaxies of stellar masses $\sim\! 10^{11}\msun$ seem to be forming a large
fraction of their stars at a star-formation rate SFR$\!\sim\!100$-$200\sy$ 
within $\sim\! 0.5\Gyr$.
This is much shorter than the age of the universe then, indicating  
coherent star formation from $\sim\! 10^{11}\msun$ of gas 
in the latest stages of its collapse/assembly rather than in its earlier 
smaller progenitors. Furthermore, the burst duration is significantly 
shorter than the time implied by the typical cosmological accretion rate 
onto haloes of $\sim\! 10^{12}\msun$ at the relevant epochs, and is
as short as the time for streaming at the virial velocity from the virial 
radius to the center. Gaseous
major {\it mergers}, which could have provided such rapid bursts, are unlikely
in many of these cases, where the galaxies are detected to be 
relatively smooth, rotating, thick discs that could not have survived a major 
merger and show no trace of such an event.
At the low end of the SFR distribution, most galaxies of stellar mass 
$\Ms \geq 10^{11}\msun$ tend to be quenched red \& dead galaxies, 
and a substantial fraction of such galaxies show strongly 
suppressed star-formation rates (SFR) already at high redshifts, $0.5<z<2.7$
\citep{kriek06,noeske07a}.
We seek an explanation for these phenomena at the two extreme regimes of SFR.

We report here on new insight gained from spherical modeling
of gas accretion onto massive galaxies, which 
could be the basis for understanding the above phenomena and
other central issues in galaxy formation.

Following the classic work on the interplay between cooling
and dynamical times in galaxy formation  
\citep{ro77,silk77,binney77,wr78,blum84},
we used analytic calculations and hydrodynamical simulations
to study the evolution of a virial shock in a spherical configuration
\citep{bd03,db06}. We predicted  
the existence of a threshold halo mass for the presence of a 
virial shock, at $\Msh\! \sim\! 10^{12}\msun$, roughly independent of redshift. 
Less massive haloes do not permit a stable shock,
as rapid radiative cooling prevents the hypothetical
post-shock gas pressure from supporting the shock against global
gravitational collapse into the halo center.
The gas accreting through the virial radius of such haloes flows
cold ($\sim\! 10^4$K) into the inner halo, where it may eventually shock,
feed a disc and efficiently form stars.
Once the halo grows above $\Msh$, a stable shock emerges from the
inner halo and rapidly propagates toward the virial radius,
halting the infalling gas and creating a hot medium in quasi-static 
equilibrium at the halo virial temperature.
We found that 
the transition to stability occurs when the standard radiative cooling time
at the assumed metallicity equals the time for compression behind the shock,
$t_{\rm comp} = (21/5) \rho/\dot{\rho} \simeq (4/3) R/V$, 
where $\rho$ is the gas density behind the shock,
$R$ is the shock radius and $V$ is the infall velocity into the shock.

Based on these findings,
it has been proposed that the presence of hot, dilute gas allows the 
suppression of gas supply to the disc, possibly assisting AGN feedback,
and thus leads to quenching of star formation \citep{binney04,db06}.
This mass threshold has proven to be the key for 
understanding the robust division of galaxies into blue, star-forming
discs versus red \& dead spheroids 
and their basic properties \citep{db06,cattaneo06,croton06}.

\begin{figure}
\vskip 8.0cm
\includegraphics{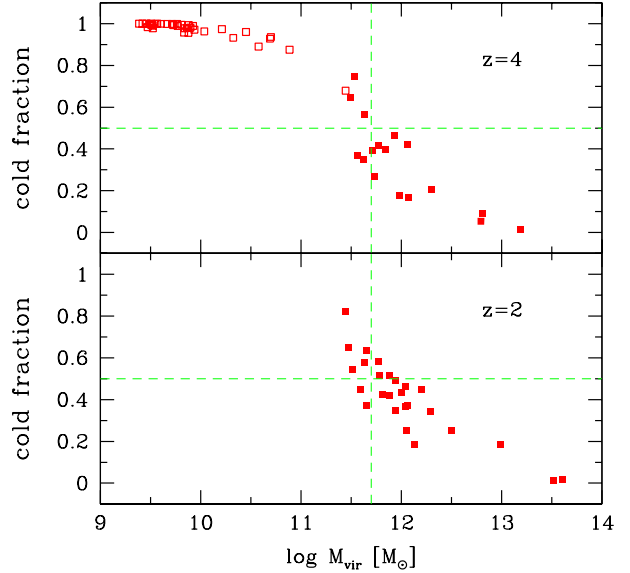}
\caption{The threshold halo mass for virial shock heating in cosmological
hydro simulations, at different redshifts, based on \citep{bdkz07}.
Each point represents the fraction of cold gas
within $0.1\Rv$-$\Rv$ in a different halo, for haloes above a minimum mass
due to numerical resolution.
The smaller haloes (open symbols) are drawn from a high-resolution
simulation in a box of comoving size $6\hmpc$, which was terminated at $z=4$
and therefore do not show counterparts at later redshifts.
The more massive haloes (solid symbols) come from a simulation of a
lower resolution in a box of $80\hmpc$.
Cold gas is defined here by $T< 10^4 + 0.4 (\Tv - 10^4)$,
e.g., $T<2.5\times 10^5$K for $\Mv \simeq 10^{12}\msun$, where
$\Tv \simeq 6\times 10^6$K.
The transition from cold dominance to hot dominance occurs near
$5\times 10^{11}\msun$ at all redshifts (vertical line), as predicted by the
spherical model for similar metallicities \citep{db06}.}
\label{fig:coldfrac}
\end{figure}

The success of the spherical model in predicting the threshold mass 
as seen in cosmological hydrodynamical simulations \citep{keres05,db06} 
is demonstrated in \fig{coldfrac}, 
which summarizes the results of \citet{bdkz07}. 
It shows as a function of halo mass the fraction of cold gas within the 
virial radius and outside the disc in dark-matter haloes from cosmological 
simulations by A.~Kravtsov \citep[described in][where they were used for other
purposes]{krav03,krav05}. 
At all redshifts, there is a transition from cold-dominated haloes to 
hot-dominated haloes near $5\times 10^{11}\msun$. This is compatible
with the threshold mass predicted by \citet{db06} for realistic halo 
metallicities that gradually grow to $Z \simeq 0.1$ solar today.  

This success of the spherical model motivates our present, more detailed 
and daring investigation of the spherical configuration. 
We study how the development of the virial shock
converts an otherwise uniform accretion through the virial radius
into a rapidly varying accretion rate onto the inner disc.
The resultant disc buildup, and the induced SFR, evolve through a 
generic sequence of events involving two quenching episodes and a massive
burst, which always starts at the crossing of the halo threshold 
mass, namely in galaxies of $\Ms\! \sim\! 10^{11}\msun$ at any redshift.
We term this phenomenon SAMBA, for Shocked-Accretion Massive
Burst and Shutdown. 

In \se{sim} we describe the simulation method.
In \se{samba} we present the four phases of the SAMBA phenomenon.
In \se{obs} we dare a tentative comparison to observations.
In \se{sam} we describe SAMBA recipes for semi-analytic modeling.
In \se{conc} we discuss our results.
In \se{app_acc}, related to \se{sim}, we provide an EPS estimate of 
the average virial accretion rate onto dark haloes.
In \se{app_sam}, related to \se{sam}, we specify the SAMBA recipes for SAMs.

\section{Method of Simulations}
\label{sec:sim}

\subsection{The Spherical hydrodynamical code}

Our accurate 1-D Lagrangian hydrodynamical code 
\citep[based on the code described in][]{bd03} 
simulates a spherical gravitating system consisting of dark 
matter and a fraction $f_b$ of gas with a constant metallicity $Z$.
The initial smooth density-perturbation profile is constructed such that it
produces the desired average accretion rate at the virial radius, \se{acc}.
The dissipationless dark-matter shells detach from
the cosmological expansion, collapse and oscillate into virial equilibrium
such that they deepen the potential well attracting the dissipating gas shells.
The gas is cooling radiatively based on the atomic cooling function computed
by \citet{sutherland93}, and contracts dissipatively into the inner halo.
The collapse of each gas shell is stopped at $\sim\! 5\%$ of the virial radius
by an artificial centrifugal force which mimics the formation of a central
disc. 

The main improvements to the original code described in \citet{bd03} are as
follows.  The numerical scheme for time evolution now computes the
pressure and energy of the gas shells within the forth-order Runge Kutta 
solution rather than externally.
A force corresponding to the cosmological constant
$\Lambda$ has been added to the force equations, making the calculation 
fully consistent with the $\Lambda$CDM cosmology, though with only minor 
effects on the evolution inside the haloes.
The simulations described here consist of $1,000$ gas shells and $5,000$ 
dark-matter shells, corresponding to a spatial resolution of $\sim\!50\pc$
in the inner halo growing to $\sim\!1\kpc$ near the virial radius.
The energy conservation level, set by the timesteps and spatial resolution,
is better than $2\%$ over today's Hubble's time. 
Convergence tests of some of the
runs with twice the spatial resolution yielded very similar results.
The baryonic fraction $\fb$ is assumed to be $10\%$ throughout the 
simulations presented here, crudely taking into account certain gas 
mass loss due to feedback effects. Simulations with $f_b=0.05\%$ and $0.15\%$ 
yield consistent results.

\subsection{Cosmological accretion rate}
\label{sec:acc}

The average fractional accretion
rate onto haloes of mass $\Mv$ at time $t$ is estimated in \se{app_acc}
using the EPS formalism, following \citet{neistein06}. 
\Equ{app_acc} reads
\be
\frac{\Mvd}{\Mv}(\Mv,t)= s(M) \frac{\delta_c}{D} \frac{\dot{D}}{D} \,,
\label{eq:acc}
\ee
\be
s(M)\equiv\left[ \frac{2/\pi}{\sigma^2(M/q)-\sigma^2(M)} \right]^{1/2} \,.
\label{eq:sm}
\ee
Here $\sigma(M)$ is the rms linear density fluctuation on the scale
corresponding to mass $M$, linearly extrapolated to $z=0$
using the linear growth function $D(t)$ normalized to unity at $t_0$. 
The constants are $\delta_c \simeq 1.68$ from the spherical collapse model,
and $q \simeq 2.2$ reflecting an intrinsic uncertainty in the EPS formalism.
The accretion time $(\Mvd/\Mv)^{-1}$ is shown later on in \fig{times}
for haloes of $10^{12}\msun$ in the $\Lambda$CDM cosmology.
A practical approximation for haloes of $\sim\! 10^{12}\msun$ in $\Lambda$CDM
is
\be
\frac{\Mvd}{\Mv}(\Mv,t)= 
0.04\,\Gyr\,\left(\frac{\Mv}{10^{12}\msun}\right)^{0.15} (1+z)^{2.25} \,.
\label{eq:acc_approx}
\ee

The mass growth history of the main progenitor of a halo that ends up with
mass $M_0$ at time $t_0$ is given by \equ{app_mph}:
\be 
M(t|M_0) = F^{-1}[\delta_c(D^{-1}-1)\, +F(M_0)] \,, 
\label{eq:mph}
\ee
where $F^{-1}$ is the inverse function of
\be
F(M) = \int_M^\infty \frac{\dd m}{m\,s(m)} \,.
\ee
We specify in \se{app_acc} the details of this calculation in $\Lambda$CDM.

The initial conditions of the simulation specify the mean density profile
of the spherical perturbation $\bar\delta_i(M)$,
referring to spheres of mass $M$ at an initial time $t_i$ in the linear regime. 
In order to end up with a halo mass $M_0$ at $t_0$, and have the proper
average virial accretion rate throughout the halo history, we invert
$M(t)$ from \equ{mph} to a collapse time for mass $M$, $t_c(M)$,
and then derive the initial profile from
\be
\bar\delta_i(M) = \delta_c \frac{D(t_i)}{D[t_c(M)]} \ .
\label{eq:deltai}
\ee

\section{Four Phases of Accretion}
\label{sec:samba}

\subsection{Generic SAMBA}

\begin{figure}
\vskip 10.2cm
\includegraphics{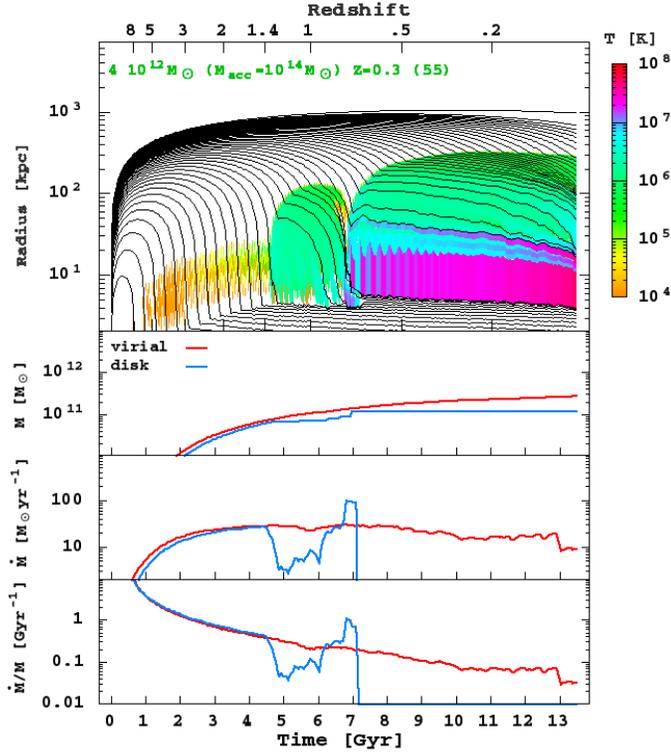}
\caption{Spherical accretion onto the outer dark halo and the inner ``disc".
{\bf Top}: Infalling gas shells, with the temperature highlighting the 
evolution of the virial shock. 
{\bf Bottom:} baryonic mass growth, accretion rate and specific accretion rate
(smoothed over a virial crossing time),
at the virial radius (red curve) and at the ``disc" (blue curve).
A {\bf late} SAMBA: today's halo mass $M_0=4\times 10^{12}\msun$,   
metallicity $Z=0.3$ and a reduced accretion rate, leading to a tentative
quenching onset at $z_1 \simeq 1.1$, a burst at $\zb \simeq 0.7$
with efficiency $B \simeq 5$, followed by very effective long-term quenching.}
\label{fig:sf55}
\end{figure}

\begin{figure}
\vskip 10.2cm
\includegraphics{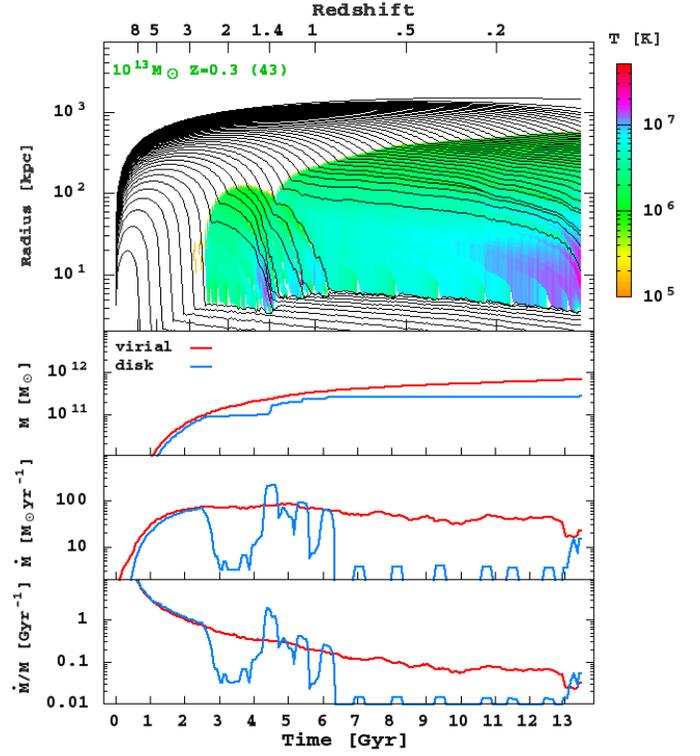}
\caption{Same as \fig{sf55} for an {\bf intermediate} SAMBA:
$M_0=10^{13}\msun$, $Z=0.3$, typical accretion rate, $z_1 \simeq 2.5$,
$\zb\simeq 1.4$ with $B\simeq 6$, followed by effective long-term quenching.} 
\label{fig:sf43}
\end{figure}

\begin{figure}
\vskip 10.2cm
\includegraphics{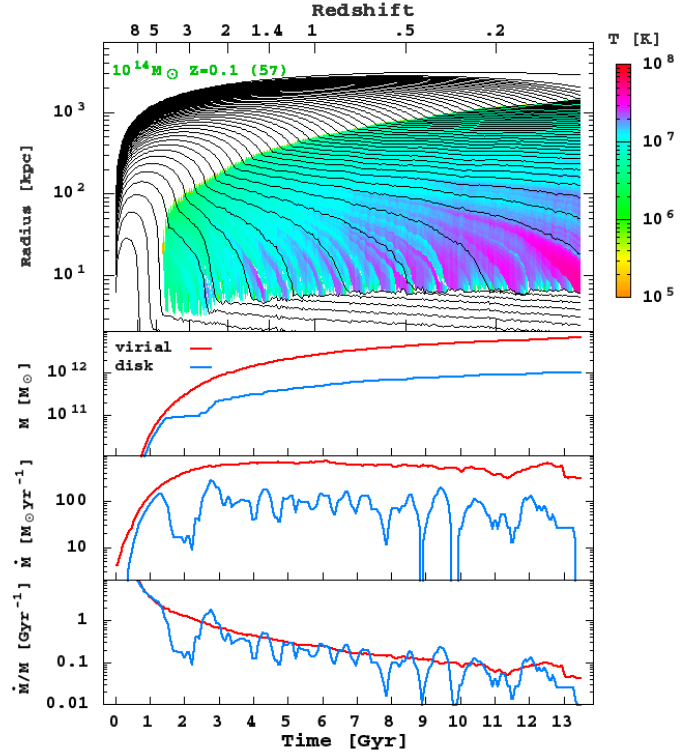}
\caption{Same as \fig{sf55} for an {\bf early} SAMBA:
$M_0=10^{14}\msun$, $Z=0.1$, typical accretion rate, $z_1 \simeq 4.8$,        
$\zb \simeq 2.5$ with $B\simeq 2.3$, followed by ineffective quenching.}
\label{fig:sf57}
\end{figure}

\Figs{sf55}, \ref{fig:sf43} and \ref{fig:sf57}
present the SAMBA phenomenon
in three simulations, spanning a range of final halo masses.
They display the radii of gas shells as they fall through the virial radius
and eventually accrete onto the inner disc. One may tentatively and very
crudely identify the disc buildup rate with maximum SFR.
We see that every SAMBA consists of four distinct phases, as follows:

\smallskip\noindent
1. {\bf Cold accretion}.
As long as the halo mass is below $\Msh$ there is no stable shock in the halo.
The gas flows cold and unperturbed toward the disc following the
cosmological accretion rate at the virial radius,
which is constructed here to be rather uniform.
The resultant disc accretion rate is growing gradually with time,
while the specific $\Msd/\Ms$ is declining.
For example, in the late SAMBA shown in \fig{sf55}, the cold accretion
phase lasts till $t \simeq 4.6\Gyr$, with the temperature map showing
no heating and the disc accretion rate rather uniform.

\smallskip\noindent
2. {\bf Tentative quenching}.
A shock forms near the disc upon crossing $\Msh$,
and it propagates outward very rapidly. The gas that falls through
the shock during its rapid expansion is heated and halted to
a stop (sometimes even pulled back) before it cools and rains down.
A first phase of suppressed disc accretion lasts for 1-2\Gyr,
e.g. between $t\simeq 2.7$ and $4.3\Gyr$ in \fig{sf43}.

\smallskip\noindent
3. {\bf Massive burst}.
The gas that was slowed down in the previous quenching
phase now joins new accreting gas in a big crunch onto the disc.
The accumulation of shells in a massive crunch results from the
longer delay suffered by gas that falls in through the shock earlier,
when the shock expands faster.
The crunch is enhanced further when the shock develops a temporary instability
causing it to contract back to sub-virial radii.
Thus, $\sim\! 10^{11}\msun$ of gas enter the disc during $\sim\! 0.5$-$1\Gyr$,
producing a burst of $\Msd\! \sim\! 100-200 \msun \yr^{-1}$,
or $\Msd/\Ms\! \sim\! 1-2$, typically twice the specific virial accretion rate.

\smallskip\noindent
4. {\bf Post-burst slowdown or shutdown}.
The halo mass is safely above the threshold, with the cooling rate
significantly slower than the compression rate, allowing the shock to
join the gradual expansion of the virial radius.
Bursts that occur prior to $z\! \sim\! 1.4$ are followed by a long period of
gradually declining specific accretion rate,
which is comparable to the specific virial accretion rate.
Bursts that occur after $z\!\sim\! 1.4$ are followed by an effective long-term
shutdown from $z\! \sim\! 0.7$ and on, lasting for several \Gyr.
The infall is halted effectively while crossing the shock during
its second expansion phase.

The timing of the SAMBA events scales with the onset time, $t_1$.
A relevant characteristic time scale at that epoch is
the virial crossing time, the time it takes to
stream at the virial velocity from the halo virial radius to the center.
Based on the spherical collapse model, it is independent of mass and given
by\footnote{
For a proper virial radius, $(1+z)^{-1}$ should be replaced by
$(\Delta_{200} \Omega_{m 0.3} h_{0.7}^2)^{-1/3} (1+z)^{-1}$.
In a flat universe $\Delta(a)\simeq 
[18\pi^2-82\Omega_\Lambda(a)-39\Omega_\lambda(a)^2]/\Omega_m(a)$,
with $a=(1+z)^{-1}$, $\Omega_m(a)$ given in \se{app_acc}
and $\Omega_\Lambda=1-\Omega_m$.}
\be
\tv \equiv \frac{R_{\rm v}}{V_{\rm v}}
\simeq 2.49 \Gyr (1+z)^{-3/2} \ .
\label{eq:tv}
\ee
In the EdS phase of cosmological evolution (valid to a good approximation
prior to $z\!\sim\! 1$), $\tv \simeq 0.18 t_{\rm Hubble}$.

We define the efficiency of disc accretion by the ratio
of specific accretion rate into the disc and into the virial radius,
\be
B\equiv \frac{\Msd/\Ms}{\Mvd/\Mv} \ .
\ee
A smoothed version of this quantity over a virial crossing time, $\Bv$,
which can be read as the ratio of the two curves shown at the bottom panels of
\figs{sf55}-\ref{fig:sf57}, 
serves us for an automatic identification of the four phases. 
This quantity is about unity during the cold-flow phase. 
We mark by $t_1$ the time when $\Bv$ starts its first drop below
unity due to the shock onset. 
Then $t_2$ and $t_3$ mark the subsequent crossings of unity upward
and downward, which we define as the beginning and the end of the burst.
Finally, we identify the following upward crossing of unity, at $t_4$,
as the end of the long-term quenching phase, when such a phase exists.

There are several robust SAMBA characteristics
that only weakly depend on the redshift of the event.
The onset of the SAMBA is always at the crossing of the threshold halo 
mass, $\Msh\! \sim\! 10^{12}\msun$, thus involving a stellar mass 
$\Ms\! \sim\! 10^{11}\msun$.
The burst peaks about half a Hubble time ($1.5-2.5\Gyr$) later, 
and involves $\lsim 10^{11}\msun$ of rapidly accreting gas in about a 
virial crossing time $\tv$ ($\sim\! 0.5$-$1\Gyr$).
The average burst efficiency is $\Bv \simeq 2$,
with a peak value of $B\!\sim\! 6$.

\subsection{Systematic redshift dependence}

By simulating the accretion histories of haloes that end up 
with different masses $M_0$ 
today, we obtain a sequence of SAMBAs with a range of starting times, $t_1$,
given that the SAMBA onset is determined by the crossing of a fixed halo 
mass threshold. Our fiducial case assumes the average virial
accretion history onto haloes $M_0$ and a constant metallicity $Z=0.3$ solar;
we obtain somewhat earlier (later) onset times by lowering (raising) $Z$
or by raising (lowering) the virial accretion rate at the relevant epoch.
It turns out that
our current fiducial choice of $Z=0.3$ at all times does not follow the gradual
cosmological growth of metallicity with time as estimated in \citet{db06}.
This results in an artificial overestimate of $\Msh$ at high $z$, by a factor
$\sim\! 2$, which we should scale out when considering the cosmological redshift
dependence of the SAMBAs. 

\begin{figure}
\vskip 8.1cm
\includegraphics{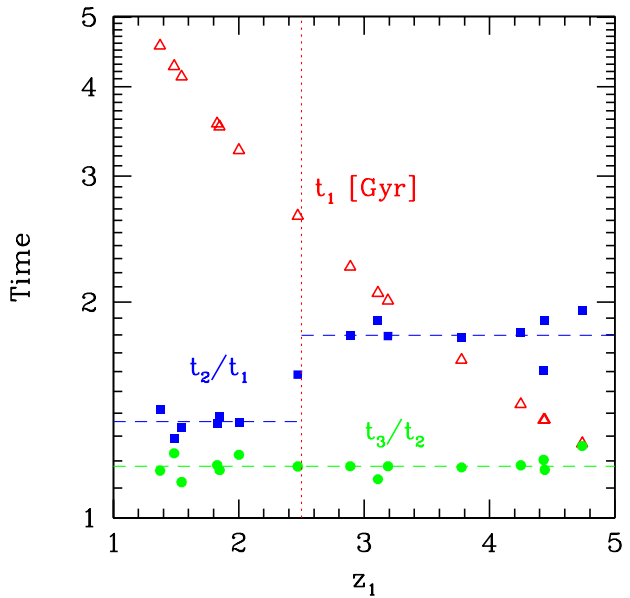}
\caption{Characteristic times of simulated SAMBAs as a function of the onset
redshift $z_1$, when $\Mv=\Msh$.
The Hubble time at the onset of tentative quenching is $t_1$
(open red triangles).
The ratio $t_2/t_1$ (filled blue squares)
refers to the duration of the tentative quenching.
The ratio $t_3/t_2$ (filled green circles) refers to the duration of the burst,
with $t_3$ the beginning of long-term quenching in late SAMBAs.
The vertical line marks the transition from early to late SAMBAs.
The horizontal lines mark the fits used for the SAM recipe in \se{sam}.}
\label{fig:sumtl}
\end{figure}

The characteristic times of 15 simulated SAMBAs are shown in \fig{sumtl}.
These times naturally scale with the Hubble time.
The burst duration is always $\Delta \tb \simeq 0.18 t_2 \simeq \tv$.
The duration of the tentative quenching is $\Delta \tq \simeq 0.8 t_1$
for $z_1>2.5$ and $\Delta \tq \simeq 0.36 t_1$ for $z_1 \leq 2.5$.
The systematic variations with the onset time are primarily driven
by the gradual decline with time of the specific virial 
accretion rate, which induces a drastic 
change between SAMBAs that start prior to $z\simeq 2.5$ and those that 
start later. This is a distinction between the central galaxies of today's
clusters of $>10^{13}\msun$ haloes and the more isolated galaxies 
in today's groups of $\leq 10^{13}\msun$ haloes. The former burst before 
$z\! \sim\! 1.4$, when $\Mvd/\Mv >1 \Gyr^{-1}$, while the latter burst later, 
when $\dot{\Mv}/\Mv <1 \Gyr^{-1}$.
In the late SAMBAs, the rapidly expanding shock overshoots  
to outside the virial radius, where the low accretion rate fails
to provide enough post-shock compression for support
against contraction. The resultant temporary instability makes the shock
drop to sub-virial radii before the pressure behind it
can push it back to the virial radius. Apparently, this instability 
appears once the specific virial accretion rate drops below $1\Gyr^{-1}$. 

\begin{figure}
\vskip 9.69cm
\includegraphics{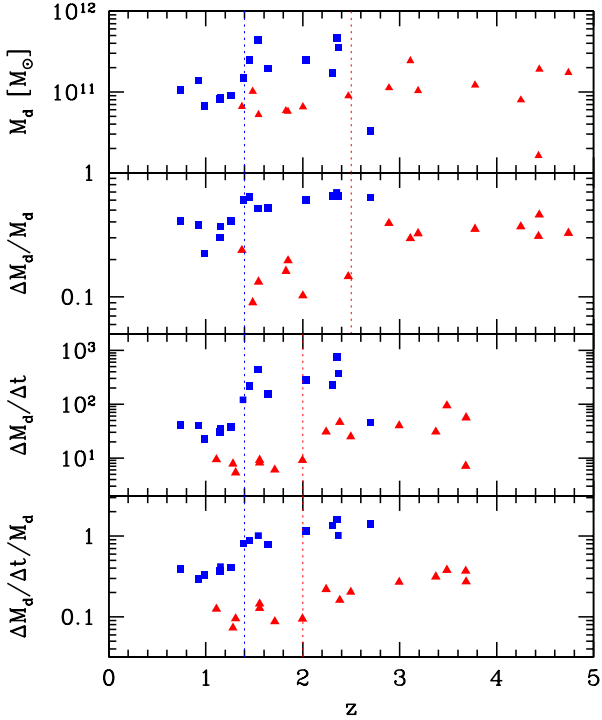}
\caption{SAMBA properties as a function of redshift: disc
accretion properties. Red triangles refer to
the tentative quenching, at $z_1$ (top panels) or
$\zq\equiv 0.5(z_1+z_2)$ (bottom panels).
Blue squares refer to the burst at $\zb\equiv 0.5(z_2+z_3)$.
Vertical lines separate early from late SAMBAs.
From top to bottom: disc mass, fractional disc growth, accretion rate,
and specific accretion rate.}
\label{fig:sumd}
\end{figure}

The effect of the virial accretion rate on the tentative shock instability 
is clearly demonstrated in \figs{sf55}-\ref{fig:sf57}. 
The strength of the instability as measured by the tentative shrinkage 
of the shock to sub-virial radii, as seen in the top panels, 
is correlated with the specific virial accretion rate at shock onset,
which could be read from the red curve at the bottom panels
or from the inverse of the accretion time shown in \fig{times}.
This trend is valid for the whole suite of simulated SAMBAs.

\begin{figure}
\vskip 9.69cm
\includegraphics{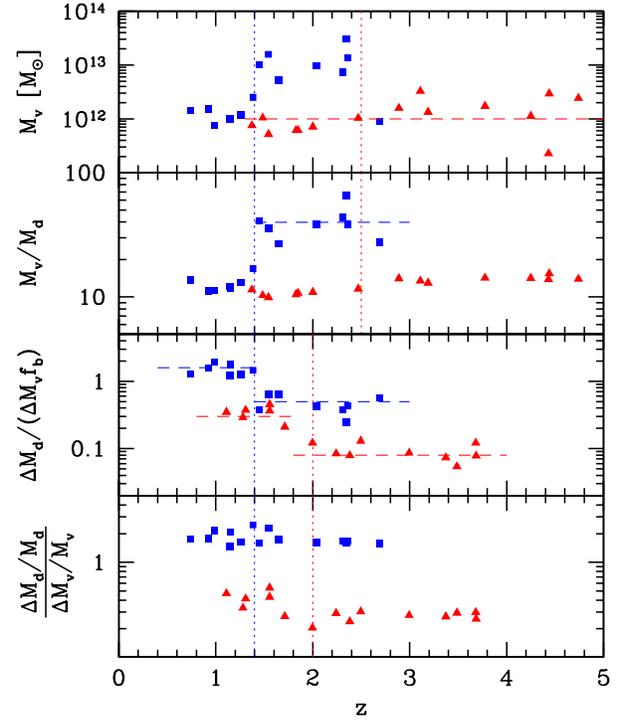}
\caption{SAMBA properties as in \fig{sumd}:
disc versus virial accretion.
From top to bottom: virial mass, virial-to-disc mass ratio,
disc-to-virial accretion ratio and efficiency $B$ of 
specific-accretion ratio.
Horizontal lines mark the fits used for SAM recipes in \se{sam}.
}
\label{fig:sumv}
\end{figure}

Other quantities characterizing the tentative quenching and the subsequent 
burst are summarized in \figs{sumd} and \ref{fig:sumv}. 
Most of the {\it burst\,} characteristics vary gradually with the burst time 
$\tb\equiv 0.5(t_2+t_3)$, 
and show a bimodality to early and late bursts separated at $z \simeq 1.4$,
associated with the shock instability.
While in the late bursts the disc mass is $\Ms\! \sim\! 10^{11}\msun$,
it is typically twice as large in the early burst, where the specific
accretion rate is higher and the preceding quenching is less efficient.
The fraction of disc mass involved in the burst between $t_2$ and $t_3$, 
$\Delta\Ms/\Ms$, varies from $\sim\! 0.7$ in the early bursts
to $\sim\! 0.3$-$0.4$ in the late ones.
The mean specific accretion rate, $(\Delta\Ms/\Delta t)/\Ms$, varies
from $\sim\! 1$-$1.5$ to $\sim\! 0.4\Gyr^{-1}$. 
This corresponds to a drop of the average accretion rate across the burst from
$\Msd\! \sim\! 100$-$400\sy$ in the early bursts to 
$\sim\! 30$-$40\sy$ in the late bursts.
The average burst efficiency over the burst is always 
$\Bb=(\Delta\Ms/\Ms)/(\Delta\Mv/\Mv) \simeq 2$, 
while the peak efficiency changes from $\sim\! 2$ in the early bursts 
to $3$-$6$ in the late bursts.

The ratio of virial to disc mass at the end of the burst is bi-modal:
It is high for the early bursts, $\Mv/\Ms\! \sim\! 25$-$50$, and low for
the late bursts, $\Mv/\Ms\! \sim\! 10$-$15$. 
The high values in the early bursts is because the high virial accretion 
rate there leads
to a rapid halo growth during the quenching phase combined with the failure
of the subsequent burst to consume most of the accumulated hot gas.
In the late SAMBAs, on the other hand, the halo growth is slow and the
burst consumes all the available gas, leading to a low $\Mv/\Ms$. 
The early bursts are thus predicted to reside in {\it very massive haloes}, of 
$\Mv \lsim 10^{13}\msun$, while the late bursts are to be found
in $\sim\! 10^{12}\msun$ haloes. 

The characteristics of the {\it tentative quenching\,} phase also
vary gradually with $t_1$.
At the onset of all SAMBAs, $\Mv\! \sim\! 10^{12}\msun$ and 
$\Ms\! \sim\! 10^{11}\msun$, 
and we should scale out the apparent slight decline with time
of these masses, which is largely an artifact of the assumed constancy of 
metallicity with time.
The typical quenching efficiency is $B\! \sim\! 0.3$ for the early SAMBAs
and $B\!\sim\! 0.4$ for the late ones. 
The ratio of disc to virial gas accretion rate, $\Delta\Ms/(\Delta\Mv \fb)$
varies from $\lsim 0.1$ to $\sim\! 0.3$ in the early and late SAMBAs 
respectively, separated at onset redshift $z \simeq 2.5$.
The tentative quenching can make the galaxy appear red for $\sim\! 1\Gyr$ 
before bursting.

The most pronounced effect of the shock instability occurring in the
late SAMBAs is a drastic {\it shutdown\,} of disc accretion for several $\Gyr$
after the burst, starting at $z\simeq 1.2$ or later. 
The infalling gas is halted and pushed back 
after crossing the shock during its second rapid expansion phase,
and the fact that the halo is already well above $\Msh$ makes the cooling
and infall time long.
These galaxies, in haloes $\leq 10^{13}\msun$ today, will appear red \& dead 
at $z \leq 1$, without the help of any additional feedback effect.
On the other hand, the post-burst phase of the early SAMBAs is 
characterized by a disc specific accretion rate that fluctuates 
about the virial rate, $B\! \sim\! 1$.
For these central cluster galaxies today to become red \& dead 
one should appeal to another quenching mechanism that could provide 
long-term maintenance.

\section{Comparison with Observations}
\label{sec:obs}

\subsection{Maximum Bursting Disks}

Two relevant examples of observed ``maximum bursters" stand out.
First, rest-frame UV/optically-selected star forming galaxies 
at $z \gsim 2$ (termed BX and BzK).
A high-resolution adaptive-optics IR study of BzK-15504 at $z \simeq 2.4$
\citep{genzel06} reveals $\Ms \simeq 8\times 10^{10}\msun$
plus $\Mg \simeq 4\times 10^{10}\msun$ forming stars at $\sim\! 100$-$200\sy$.
The stellar population has a typical age of only $\sim\! 0.5\Gyr$,
less than 20\% of the Hubble time at that epoch.
The morphology and kinematics fit an exponential disc of
scale radius $\simeq 4.5\kpc$ rotating at $V_{\rm rot} \simeq 230\kms$,
and a recent major merger is ruled out.
The velocity dispersion of $\sigma \simeq 60-110\kms$ indicates a rapid
accretion onto the disc, again of $\sim\! 0.5\Gyr$.
Several galaxies with similar properties were detected in a
lower resolution study \citep{forster06}.

Other examples are found among the LIRGs at $z\! \sim\! 0.6$-$0.9$.
\citet{hammer05} have detected at least six galaxies of 
$\Ms\! \sim\! 10^{11}\msun$
forming stars at $\simeq 120-200\sy$ while their morphologies and kinematics
resemble spiral galaxies with no trace of major mergers.
More than half of their IR-luminous galaxies are spirals.
The compilation of SFR at $z\!\sim\! 1$ by \citet{noeske07a} reveals
some massive galaxies that form stars at $\sim\! 200\sy$,
while only $\sim\! 7\%$ of the galaxies show evidence for ongoing major
mergers \citep{lotz06}, and this fraction is not higher for the galaxies
with the highest SFR (private communication).

In summary, the observed maximum bursters show the following robust features:
(a) the baryonic mass is on the order of $\Ms\! \sim\! 10^{11}\msun$,
(b) a substantial fraction of the baryonic mass is bursting, 
(c) the burst is short compared to the Hubble time and the cosmological mean
accretion time, and is comparable to the virial crossing time,
(d) the star formation occurs in an extended configuration of a few \kpc,
and 
(e) in many cases there are morphological and kinematical indications
for a thick rotating disc with no trace of major mergers.

\begin{figure}
\vskip 8.1cm
\includegraphics{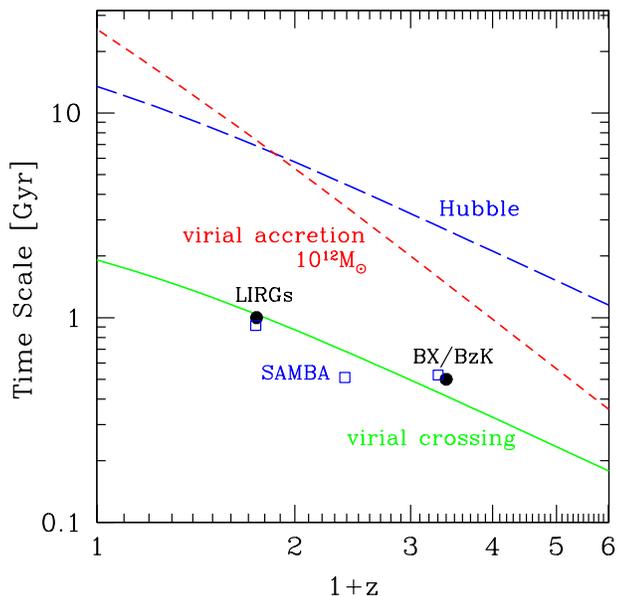}
\caption{Characteristic time scales as a function of redshift. 
The curves refer to
(a) the Hubble time (blue, long dash), 
(b) the average cosmological accretion time onto $\Mv = 10^{12}\msun$ 
haloes (red, short dash), and
(c) rapid streaming from the virial radius to the center, $\Rv/\Vv$ (green,
solid).
The estimated duration of the observed maximum starbursts are represented 
by black circles circles.
The predicted duration of three SAMBA bursts are marked by blue open squares.
}
\label{fig:times}
\end{figure}

A distinct feature of the maximum bursts is that most of the
stars of the $\Ms\! \sim\! 10^{11}\msun$ galaxies
have formed over a short period of time just prior to the observed epoch,
$\sim\! 0.5 \Gyr$ at $z \gsim 2$ and $\sim\! 1 \Gyr$ at $z \lsim 1$.
This is compared in \fig{times} to several relevant characteristic timescales.
The burst duration is shorter than the
age of the universe at that time by a factor $\sim\! 6$.
This implies that most of the stars have
formed in a coherent starburst of the whole $\sim\! 10^{11}\msun$ of gas
in its late stages of collapse or assembly rather than in the smaller
progenitors during the preceding period of hierarchical buildup.
The implied suppression of SFR in small haloes and its possible origin will
be discussed in \se{conc}.

The burst duration is also 
shorter than the characteristic time for cosmological accretion onto
dark haloes of $\sim\! 10^{12}\msun$,
the likely hosts of the $\Ms\! \sim\! 10^{11}\msun$ maximum bursters,
as given by the inverse of the average $\Mvd/\Mv$ in \equ{acc}.
At $z=0.75$, one has $t_{\rm acc}(10^{12}\msun) \simeq 7\Gyr$, 
much longer than the indicated observed duration of $\Delta t\! \sim\! 1\Gyr$.
At $z=2.4$, one has $t_{\rm acc}(10^{12}\msun) \simeq 1.6\Gyr$, 
which is about 3 times the observed $\Delta t\! \sim\! 0.5\Gyr$,
though if the relevant haloes are more massive, $\lsim 10^{13}\msun$,
the accretion time is $\sim\!50\%$ shorter.   
As seen in \fig{times}, the burst durations are rather comparable to the
shortest characteristic time of the system, $\tv$. 

Gas-rich {\it major mergers}, which could in principle produce such rapid, 
massive bursts, fail to match the robust features (a), (d) and (e). 
In particular,
major mergers are expected to destroy the discs and be associated with 
perturbed clumpy morphology and kinematics rather than a smooth rotating disc. 
This is because the dynamical disturbance, involving tidal effects
and violent relaxation, is expected to be pushed to an extreme during 
the $\sim\!100\Myr$ of the close passages.
Furthermore, the induced star formation is expected to be concentrated 
in a central cusp rather than an extended disc. 
It is worth noting that the mergers are only weakly associated 
with a characteristic mass, the nonlinear clustering scale $\Mps$, 
which varies rapidly with time --- they are not associated with galaxies
of any specific fixed mass.

The SAMBA bursts, as described by our naive spherical model
above, have the potential of reproducing the observed features one by one
in a straightforward way. In particular, they have the potential of keeping the
disc intact while doubling its mass and puffing up its thickness.
This is because the disc buildup, even if it involves filamentary and clumpy
cold flows, could be rather uniformly stretched over the 
$\sim\!0.5$-$1\Gyr$ duration of the burst, and therefore involve
only moderate perturbative effects on the dynamics. 
In addition,
the SAMBA bursts are predicted to be specifically associated
with one characteristic baryonic mass of $\sim\!10^{11}\msun$ at all redshifts.

The early SAMBA bursts, which may be associated with the observed
maximum bursters at $z \gsim 2$, are predicted to be in 
$\Ms\! \sim\! 2\times 10^{11}\msun$ discs
embedded in rather massive haloes of $\Mv\! \sim\! 6\times 10^{12}\msun$.
The comoving number density of haloes of $\Mv > 5\times 10^{12}\msun$
at $z=2.2$ is $n \simeq 10^{-4} (\hmpc)^{-3}$,
consistent with the number density of observed maximum bursters 
\citep{genzel06}\footnote{This estimate is for the WMAP3 parameters of 
the $\Lambda$CDM cosmology, $\omm=0.24$, $\oml=0.76$, $h=0.73$, and
$\sigma_8=0.74$\citep{spergel06_wmap3}. A similar number density is obtained 
for haloes twice as massive if one adopts instead $\omm=0.3$, $\oml=0.7$,
$h=0.7$, and $\sigma_8=0.9$.}
The predicted halo virial velocity is $\sim\! 400\kms$ and the virial radius is 
$\sim\! 200\kpc$. The low concentration expected for such
haloes at $z\! \sim\! 2.2$, $C\!\sim\!4$ \citep{bullock01_c}, 
implies that the circular velocity at $\sim\! 10\kpc$
should be $\sim\! 200\kms$, compatible with the observed bursters.
The large virial radius for that epoch, with an assumed typical spin parameter, 
implies an exponential disc radius of $\sim\! 5\kpc$ \citep{bullock01_j},
which explains the observed extended discs. 

Thus, the burst properties predicted by the idealized spherical model do
surprisingly well in matching the observed bursts.

\subsection{Suppressed Star Formation Rate}

\begin{figure}
\vskip 6.2cm 
\includegraphics{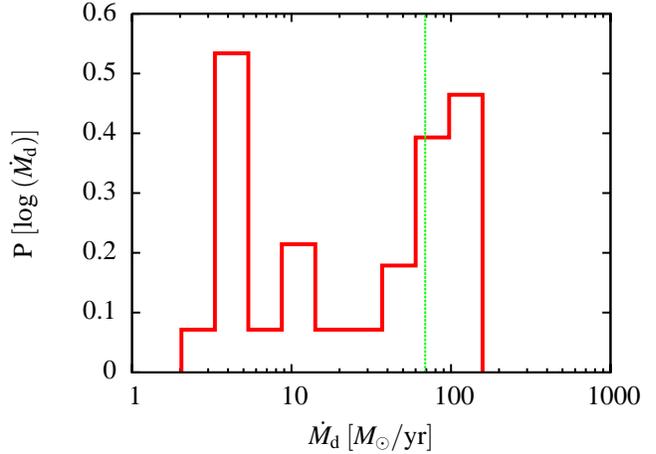}
\caption{Predicted distribution of log disc accretion rate for galaxies in 
the disc-mass bin $(0.5$-$2)\times 10^{11}\msun$ near $z\!\sim\! 2$. 
The vertical green line marks the mean virial accretion rate. 
About half the galaxies are expected to show a high accretion rate into the
disc, while the other half is predicted to suffer a strongly suppressed 
accretion rate.
}
\label{fig:sfrhist}
\end{figure}

Despite the apparent vigor of the burst event, the most robust feature of
the SAMBAs is the quenching.
We note that during the SAMBA sequence of events, from the onset of the shock
till the end of the burst about a Hubble time later, 
each galaxy spends more time in a phase of
suppressed accretion rate rather than in a rapid accretion mode.
\Fig{sfrhist} shows a crude estimate of the predicted distribution 
of disc accretion rate for galaxies in the stellar mass bin 
$(0.5$-$2)\times 10^{11}\msun$ near $z\!\sim\! 2$. This crude estimate
is based on the fraction of time spent at each accretion rate
in the simulated SAMBA shown in \fig{sf43}, assuming
that it is a representative SAMBA at that epoch.
While a fraction of the galaxies are expected to show a high accretion 
rate into the disc at about twice the average virial level, 
{\it more than half\,} the galaxies
are predicted to present a strongly suppressed disc accretion rate,
an order of magnitude below the virial accretion rate or less.
The associated distribution of SFR may be qualitatively similar.

Preliminary
indications for a substantial population of $\sim\! 10^{11}\msun$ galaxies
with strongly suppressed SFR already exist. 
A bimodality in the SFR distribution is apparent at $z \lsim 1$ 
in \citet[][Fig.~1]{noeske07a}, where the typical SFR is $>10\sy$ 
while about half the galaxies have an upper limit of SFR$<1\sy$.   
At $z\! \sim\! 2$-$2.7$, \citet{kriek06} report that a significant fraction of 
the $\sim\! 10^{11}\msun$ galaxies have surprisingly low upper limits for the
specific SFR at $< 0.1 \Gyr^{-1}$, while the typical is $\sim\! 1\Gyr^{-1}$.

The SAMBAs predict that the SFR histories of galaxies should show
two peaks of SFR separated by a couple of Gyr, but this might be
hard to detect observationally.

\subsection{Downsizing}

The SAMBA phenomenon provides a simple explanation for the observed
``downsizing" of elliptical galaxies. 
An ``archaeological" analysis of stellar ages reveals
that today's more massive ellipticals have formed their stars earlier
and over a shorter time span \citep{thomas05}.
From their SFR histories as a function of look-back time one learns that 
elliptical galaxies of present-day stellar mass $\simeq 10^{11}-10^{12}\msun$
have formed most of their stars at $z\! \sim\! 2$-$3$
in a maximum burst over $1$-$0.5\Gyr$ with $\Msd/\Ms\! \sim\! 2$
(or $z\! \sim\! 1.2$-$1.6$ in lower density environments).
The progenitors of today's ellipticals, prior to their subsequent growth
via dry mergers, seem to be consistent with the
maximum bursters seen at high redshift, and with the SAMBA predictions.

The constancy of $\Msh$ with redshift, which implies that all galaxies
burst when they are $\Ms\! \sim\! 10^{11}\msun$ and quench immediately after
automatically implies downsizing in both the peak of SFR and the subsequent
quenching. Galaxies of larger stellar mass today are likely to be embedded in 
more massive haloes, which in turn have crossed the threshold mass 
for the onset of SAMBA at an earlier time. They bursted earlier,
for a shorter duration, and shut down earlier accordingly.
This is demonstrated in a
cosmological semi-analytic simulation by \citet{cattaneo07_ds}. 

The SAMBAs predict yet another kind of downsizing phenomenon, 
concerning halo mass at burst:
the early bursts at $z > 1.4$ are predicted to reside in massive haloes  
only slightly smaller than $\Mv\!\sim\!10^{13}\msun$, 
while the late bursts are to be found in smaller haloes, 
$\Mv\! \sim\! 10^{12}\msun$.

\section{Recipes for Semi-Analytic Models}
\label{sec:sam}

In order to study the potential observable implications of the SAMBA 
phenomenon in more detail, one can translate the predictions from 
the spherical model into a recipe for gas cooling and accretion rate
onto the disc. 
This recipe is to be incorporated in Semi-Analytic Models (SAM), 
which follow the baryonic physics in given dark-halo merger trees. 
The SAMBA recipe should replace the standard recipe adopted in SAMs
\citep{white91}, which assumes that the
halo gas interior to a ``cooling radius" is accreted onto the disc.
The cooling radius within a (spherical) halo of a given mass at a given time
is traditionally determined by the position where the cooling time equals 
a certain dynamical time, e.g., the Hubble time. 
The SAMBA phenomenon suggests a deterministic modification of the disc 
accretion rate, based on our spherical modeling, with no free parameters.
However,
an optimal match to observations may suggest a certain fine-tuning of the
model parameters, compensating for deviations from spherical symmetry and
uniform accretion.

\subsection{Following Halo Evolution in SAM}

\begin{figure}
\vskip 8.5cm
\includegraphics{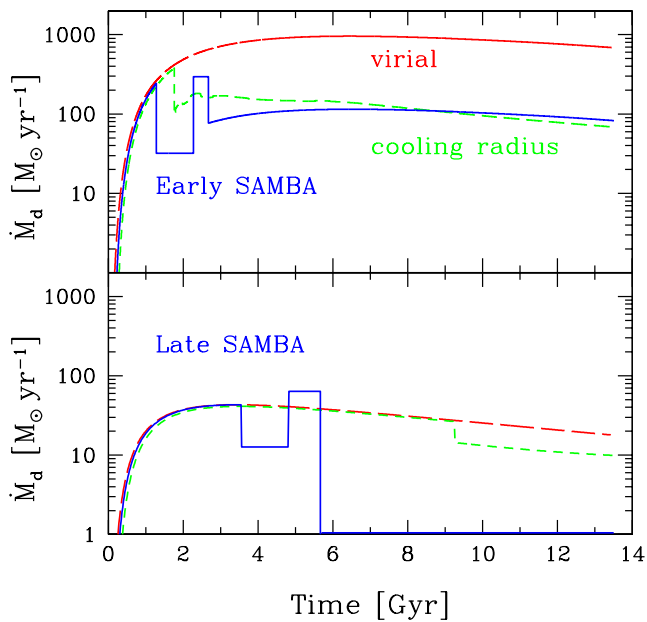}
\caption{Disc growth rate in the SAMBA recipe for SAM (blue, solid)
versus the smooth virial baryonic accretion rate (red, long dash) and the
SAM standard (green, short dash) based on the evolution of the
``cooling radius" a la
\citet{white91}. {\bf Top:} an early SAMBA ($z_1\simeq 4.8$, $\Mv_0\simeq
10^{14}\msun$). In this case, long-term quenching has to originate
from another process.
{\bf Bottom:} a late SAMBA ($z_1\simeq 1.8$,
$\Mv_0\simeq 4\times10^{12}\msun$),
were long-term quenching is a natural outcome of uniform accretion.
}
\label{fig:sam}
\end{figure}

The proposed SAMBA recipe for the accretion rate onto the disc is
specified in \se{app_sam_sam}.  In summary,
given the mass growth of each halo in the merger tree, $\Mv(t)$, 
and its derivative $\Mvd$, the SAMBA onset time $t_1$ (redshift $z_1$)
is identified by $\Mv(t_1)=M_1=10^{12}\msun$, and
the SAMBA is classified as ``early type" or ``late type" according to
whether $z_1>2.5$ or $z_1\leq 2.5$ respectively.
Then the times $t_2$ and $t_3$, the beginning and the end of the burst,
are computed using the fits shown in \fig{sumtl}.
The algorithm then specifies the growth rates of cold and hot gas mass 
during each of the four SAMBA phases defined by the above times.
The numerical factors used for the tentative 
quenching and for the burst are read from straightforward fits 
to $\Delta \Ms/(\Delta \Mv \fb)$ in \fig{sumv}.

The disc mass growth according to the proposed SAMBA recipe for SAM is
shown in \fig{sam} for typical early and late SAMBAs.
It is compared to the 
standard disc growth rate following the evolution of a ``cooling radius". 
The latter has a characteristic drop when the cooling radius first becomes
smaller than the virial radius.
The cooling-radius recipe can be somewhat improved once the Hubble time
used in its definition is replaced by $\tv = 0.18\,t_{\rm Hubble}$.
Yet, it fails to capture the depth of the tentative quenching, the 
burst, and the long-term quenching of late SAMBAs.

Recall that the early SAMBAs, ending up in $\Mv \geq 10^{13}\msun$ haloes today,
should be quenched after $t_3$ by another mechanism, such as clumpy 
accretion or AGN feedback. If such a process is not explicitly
included in the SAM,
one can simply set $\Msd=0$ in the post-burst phase for all haloes \citep[as
in][]{cattaneo06}.

\subsection{SFR as a function of Halo Mass: FHM}

\begin{figure}
\vskip 8.5cm
\includegraphics{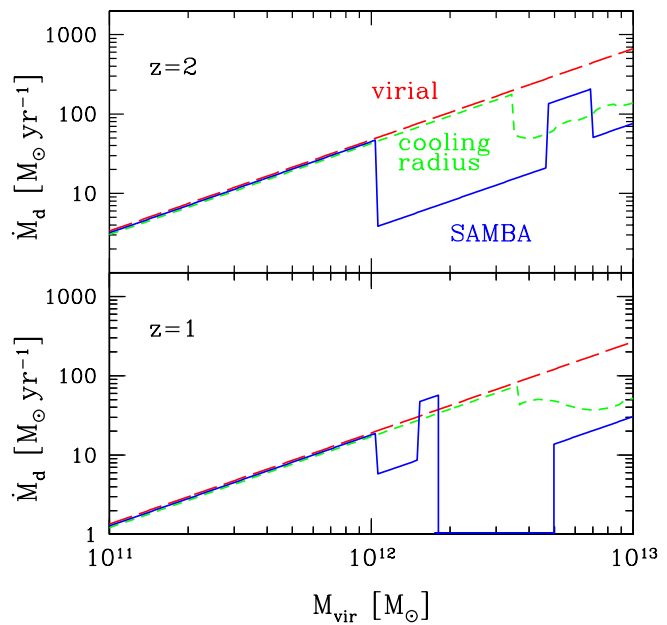}
\caption{Disc growth in the SAMBA recipe for FHM (blue, solid)
versus the virial baryonic accretion rate (red, long dash) and the standard
based on ``cooling radius" (green, short dash).
{\bf Top:} at $z=2$.  {\bf Bottom:} at $z=1$.
Quenching in haloes $\sim\!10^{13}\msun$ and larger has to be due to another
process. 
}
\label{fig:fhm}
\end{figure}

An alternative, quick and crude way to learn about the implications of
SAMBAs is by providing a recipe for the SFR as a function of halo mass
and redshift, to be incorporated in a given halo population in a partial SAM
while bypassing the detailed baryonic processes of cooling, star formation, 
and feedback \citep[Formation History Modeling, FHM,][]{croton07}.
The fact that the SAMBA phenomenon is driven by halo mass allows us to
provide a concrete recipe of this sort, specified in \se{app_sam_fhm}.

In summary, one starts with $M_1=10^{12}\msun$ at $t_1$,
and computes the other SAMBA characteristic halo masses $M_2$ and $M_3$,
corresponding to the characteristic times $t_2$ and $t_3$,
using fits to our simulation results.
Given the virial accretion rate in the provided merger tree, $\Mvd$,
the disc accretion rate, which could be interpreted as an upper limit for
the SFR, is given in each of the mass ranges, corresponding to the four
SAMBA phases.
The numerical values are read
from crude fits to the simulation results for $\Delta \Ms/(\Delta \Mv \fb)$,
as shown in \fig{sumv}. 
When implementing the recipe, one can either use 
the actual mass growth rate of the haloes drawn from merger-tree realizations,
or the average virial accretion rate as provided in \equ{acc_approx}. 

The SAMBA recipe for disc accretion rate is
shown in \fig{fhm}, compared to the standard rate based on a ``cooling radius". 
Again, the cooling-radius recipe can be somewhat improved once the Hubble time
is replaced by $\tv$, but it would still fail to recover the burst and the
deep quenching features of the SAMBAs.

The main feature of the SAMBA FHM is the quenching of SFR above a threshold
halo mass, $\Msh\! \sim\! 10^{12}\msun$, as already implemented in SAM
simulations \citep{cattaneo06,croton06}.
The interesting new feature predicted by the model is the high SFR in
a second range of masses above $\Msh$. At low redshifts, this is a
slight increase of the effective threshold mass for quenching to 
$\sim\! 2\times 10^{12}\msun$.
At $z\geq 1.3$, the high SFR is predicted to extend to haloes as large as 
$\sim\! 7\times 10^{12}\msun$. This should allow the appearance of massive blue
galaxies at high redshifts.
In the special range $1.24< z <1.53$, the high SFR is expected over an
especially broad range of masses, $1.5$-$7\times 10^{12}\msun$ (not shown in
\fig{fhm}). 
Given that $10^{12}\msun$ is the typical
mass for haloes that form at this redshift range, this should lead to
a peak in the history of cosmological SFR density (the ``Madau" plot).

\section{Discussion and Conclusion}
\label{sec:conc}

A straightforward spherical model of uniform gas accretion onto haloes
predicts a robust, rich sequence of events in the accretion onto 
the central discs, SAMBA, which could be the major driver of star-formation 
history in galaxies.
An initial {\it cold accretion} phase ends abruptly upon
the birth of a shock in the inner regions of a halo once its mass exceeds
a threshold mass of $\sim\! 10^{12}\msun$ and the disc contains $\sim\!
10^{11}\msun$. 
The infall of gas through the expanding shock creates a period of
{\it tentative quenching}, where the accretion onto the disc is 
suppressed for a couple of Gyr.
The accumulating hot gas joins new incoming gas in a big crunch,
leading to a massive, rapid, {\it burst\,} involving $\sim\! 10^{11}\msun$ 
of gas in a few hundred Myr. The subsequent expansion of the shock
to the virial radius results in a {\it long-term shutdown},
which is especially efficient if the burst occurs after $z\! \sim\! 1.4$,
corresponding to today's haloes of $\leq 10^{13}\msun$.
The predicted SAMBA bursts seem to match the observed maximum bursters of
$\Ms\! \sim\! 10^{11}\msun$ at $z\lsim 1$
and at $z\gsim 2$, as well as the indicated population of 
$\Ms\! \sim\! 10^{11}\msun$ galaxies with suppressed SFR.
This rich phenomenon is a generic consequence of uniform accretion,
before including the effects of major mergers or AGN feedback.

The implications of the SAMBA phenomenon should be taken with a grain of
salt because the relevance of the spherical uniform accretion model
to the actual SFR history in real galaxies is yet to be demonstrated. 
This is an immediate challenge for cosmological simulations.
The actual accretion could be rather filamentary and clumpy to begin with, 
and one should verify the extent to which the SAMBA behavior remains valid
under such conditions.
One could be encouraged though by
the success of the spherical model in predicting the threshold mass 
for virial shock heating \citep{keres05,db06,bdkz07}.
First clues for natural long-term quenching of the sort predicted here 
in a $\sim\! 10^{12}\msun$ halo may be seen in high-resolution 
simulations of galaxies with quiet histories \citep{naab07,libeskind07}.
The following discussion of possible implications is pursued  
under the tentative assumption that other robust SAMBA features will also
be reproduced in cosmological simulations of the formation of such galaxies. 
The apparent match of the spherical predictions with several different 
observations makes this discussion worthwhile.

Using the fact that the accretion onto the disc is a straightforward
function of halo mass, we propose a simple accretion recipe for SAMs,
to replace the standard recipe based on a ``cooling radius".
A realistic study of the SFR that is associated with this galaxy buildup,
both in the case of smooth accretion of minor mergers
and under major mergers and feedback effects,
is necessary for a reliable comparison with observations.
Still, the naive interpretation of the disc accretion rate as the maximum
possible SFR provides a preliminary indication for a surprisingly good
match to observations.

An obvious necessary condition for SAMBAs (which is actually
automatically implied
by the observed maximum bursts independent of any model)
is that a substantial fraction of 
the accreting material onto haloes of $\sim\! 10^{12}\msun$ is gaseous 
rather than stellar.  
Star formation could have been suppressed in haloes smaller than a certain
threshold due to different processes, such as stellar feedback
or a density threshold for star formation.
Given such a threshold mass, we can estimate the gas fraction in the accretion
by evaluating the mass in all accreting haloes below that threshold 
compared to the total accreting mass \citep{neistein07}.
We find using the EPS formalism specified in \se{app_acc} that
about one half of the accreting mass onto a halo of $10^{12}\msun$ is
in haloes smaller than $\sim\! 10^{11}\msun$ (or ``smooth" accretion),
independent of time.
This should therefore be the minimum halo mass for star formation 
in order to have a gas-dominated accretion.
If the suppression is due to supernova feedback, the minimum mass for
star formation corresponds to haloes of virial velocity $\sim\! 120\kms$ 
\citep[][Fig.~2]{ds86,db06}, for which the accretion is almost all gaseous
at $z=0$, it is about 90\% at $z \simeq 0.7$, and about 67\% at $z \simeq 2.4$.
This hints that supernova feedback could be the suppression mechanism
allowing the accretion onto $\sim\! \Msh$ haloes to be predominantly 
gaseous at all relevant redshifts. 

The dramatic variations in the disc accretion rate imply that each 
galaxy is likely to evolve through several subsequent phases of very 
high and very low SFR. 
This means that the expected overall trend of evolution from the blue cloud 
to the red sequence of galaxies is not a one-way track. In particular,
some of the red \& dead galaxies observed at high redshift may actually 
become blue again at later redshifts. 
The lesson is that the interpretation of the observed evolution of the 
blue and red luminosity functions may involve non-trivial book-keeping.

The late SAMBAs offer a very efficient post-burst shutdown that lasts 
for 6-7 Gyr. This can naturally give rise to red \& dead galaxies 
in today's small-group haloes of $\leq 10^{13}\msun$, perhaps S0 
or elliptical galaxies \citep[as simulated by][]{naab07,libeskind07}.  

In the early SAMBAs, the accretion rate into the disc is suppressed compared
to the virial rate even during the burst phase, but the post-burst
accretion rate is not negligible. For these central galaxies in today's cluster
haloes of $> 10^{13}\msun$ to end up as large red \& dead ellipticals 
one needs a quenching mechanism beyond the one provided by spherical 
accretion.  
We show elsewhere \citep{bd07}
that this could be provided by clumpy accretion, in which gas
clumps of $10^7-10^9\msun$ transfer their gravitational energy via drag
and local shocks into heating and puffing up of
the inner-halo gas of $\geq 10^{13}\msun$ haloes.  

Alternatively, a fashionable scenario suggests that the required quenching is
provided by AGN feedback \citep{croton06,cattaneo06}. 
It has been noticed that the shock 
heating of the gas in haloes $>10^{12}\msun$ enables the required coupling 
of the AGN energy with the halo gas \citep{db06}, but we now realize
that the SAMBA burst may be the actual trigger for AGN activity just
above this critical mass. The SAMBAs may thus be responsible for both, the
energy source and the coupling mechanism.

If the SAMBAs are responsible for triggering AGN activity, recall that 
they do so only in galaxies of $\sim\! 10^{11}\msun$, a couple of Gyr after
their haloes became larger than $\sim\! 10^{12}\msun$ . 
This may explain the otherwise surprising result that
the bright quasars (near $L_*$ of their Schechter luminosity function)
seem to reside at all redshifts in haloes
of $\sim\! 2\times 10^{12}\msun$ \citep{croom05}.
A straightforward calculation using the Press-Schechter formalism shows
that the evolution of number density of quasars at $z > 1$
can be reproduced by assuming quasar onset in every halo when it first becomes 
$\sim\! 2\times 10^{12}\msun$ and allowing a fixed short
lifetime for each quasar.

The SAMBA events offer a complementary alternative to wet major
mergers in producing rapid, massive bursts, with the important difference that
it can leave the disc intact. The associated shutdown leads to red \& dead 
galaxies in haloes above the threshold mass.
However, while major mergers also automatically produce spheroids, what 
could be the mechanism responsible for spheroid formation in association
with the old, gas-poor stellar population according the halo shutdown
scenario? 
First, the SAMBA burst provides the gas necessary for bar instability in 
the disc, which then transfers angular momentum to the outer disc
and halo and produces a bulge.
Second, the SAMBA quenching makes any subsequent mergers dryer,
so they could lead to the large, boxy, non-rotating
ellipticals, which cannot be reproduced by wet major mergers
\citep{novak07}.
Third, the minor mergers associated with the continuous smooth accretion of 
SAMBA could themselves lead to the formation of a spheroid 
\citep{naab07,bournaud07,libeskind07}.

We argued in \citet{db06} that
discs could be built and form stars by cold streams, which could persist
as narrow filaments even within the shock heated medium slightly above
$\Msh$, especially at high redshifts.
Streaming at the virial velocity, these flows could indeed give rise to
rapid starbursts, on a time scale comparable to the virial crossing time,
but it is not clear that these flows could bring in mass in a rate that is
much more efficient than the cosmological accretion rate. The SAMBAs provide
such more efficient bursts in a natural way.

All the above is yet to be confirmed in cosmological simulations.
It would be interesting to learn how the SAMBA phenomenon manifests itself when
the spherical accretion of minor mergers is accompanied by filamentary streams
and major mergers.

\section*{Acknowledgments}
We acknowledge stimulating discussions with 
S.M.~Faber, R.~Genzel, F.~Hammer \& J.P.~Ostriker.
This research has been supported by ISF 213/02 and GIF I-895-207.7/2005.

\bibliographystyle{mn2e}
\bibliography{dekel}

\appendix
\section{Main Progenitor History}
\label{sec:app_acc}

Following \citet{neistein06}, we use the EPS formalism\citep{lacey93} to
derive an analytic estimate for the main progenitor history.
We track time backwards via the variable 
$\omega=\delta_c/D(t)$, 
where $D(t)$ is the cosmological growth factor of linear density fluctuations
normalized to $D(t_0)=1$ (see below), and $\delta_c\simeq1.68$ 
(ignoring small variations due to the actual cosmology used).
Mass enters via the monotonically decreasing function $\sigma(M)$,
the standard deviation of the initial density fluctuations, as derived from
their power spectrum, smoothed on a scale that corresponds to mass $M$ 
and linearly extrapolated to $t_0$ (see below). 
According to EPS, a halo of mass $M_0$ at $\omega_0$
has the following average number of progenitors of mass
in the range $(M,M+dM)$ at the earlier time $\omega_0+\Delta\omega$:
\bea
\dd N\!=\!\frac{1}{\sqrt{2\pi}}  
\frac{\Delta\omega}{(\Delta\sigma^2)^{3/2}}  
\exp\! \left[-\frac{(\Delta\omega)^2}{2 \Delta \sigma^2}\right]  
\left\vert {{\dd}\sigma^2 \over {{\dd}M}} \right\vert  
\frac{M_0}{M}\,
{\dd}M ,
\label{eq:EPS} 
\eea
where $\Delta\omega=\omega-\omega_0$ and
$\Delta \sigma^2=\sigma^2(M)-\sigma^2(M_0)$.

\subsection{Accretion Rate onto $M$ at $t$: a Small Time Step}

We wish to compute the main progenitor history, i.e., the conditional average
$M(\omega|\omega_0,M_0)$.
The main progenitor is guaranteed to also be the most massive progenitor
when $\Delta \omega$ is small enough, and this allows us to compute
$\dot{M}$ by evaluating $M(\omega_0+\Delta\omega|\omega_0,M_0)$
for small $\Delta\omega$.

For $M\geq M_0/2$, the probability density $P(M,\omega |M_0,\omega_0)$
is the same as the total progenitor distribution $\dd
N/\dd M$ given by equation \ref{eq:EPS}. For $M < M_0/2$,
however, the only condition is $P \leq \dd N/\dd M$, and it is not 
sufficient to constrain $P$.  As a first approximation
we assume that the main progenitor always has a mass $M\geq M_0/2$.
Then, the average mass of the main progenitor is
\bea
M(\omega_0+\Delta\omega |\omega_0,M_0)  = 
\int_{M_0/2} ^{M_0} {{\dd}N \over {\dd}M} M \dd M \\
\nonumber  = 
M_0 \left[ 1 - 
{\rm erf}\left( \frac{\dW}{\sqrt{2\sigma^2_q-2\sigma^2_0}}\right) \right] \,,
\label{eq:mmainaver} 
\eea
where $\sigma_q=\sigma(M_0/q)$ with $q=2$, and $\sigma_0=\sigma(M_0)$.
The rate of change is then
\bea
\frac{\dd M }{\dd\omega} & = & 
\lim_{\dW \rightarrow 0} \frac{ M(\omega_0+\dW|\omega_0,M_0) - M_0 }{\dW} \\ 
\nonumber & = & -M_0 \lim_{\dW \rightarrow 0} \frac{1}{\dW} \rm{erf}\left(
\frac{\dW}{\sqrt{2\sigma^2_q-2\sigma^2_0}} \right) \,.
\eea
Using the limit $\rm{erf}(x)  \rightarrow 2x/\sqrt{\pi}$ when
$x\rightarrow 0$ we get
\bea
\label{eq:m_asm} 
\frac{1}{M}\frac{\dd M}{\dd\omega} =
-\left(\frac{2/\pi}{\sigma^2_q-\sigma^2} \right)^{1/2} \,.
\eea
Thus, the mean specific accretion rate onto haloes of mass $M$ at $t$ is
\be
\frac{\dot{M}}{M}\left(M,t\right) = -s(M)\, \dot\omega(t) \,,
\label{eq:app_acc}
\ee
with
\be 
s(M)\equiv\left[ \frac{2/\pi}{\sigma^2(M/q)-\sigma^2(M)} \right]^{1/2} \,,
\label{eq:app_sm}
\ee
\be
\dot\omega(t) = -\frac{\delta_c}{D}\frac{\dot{D}}{D} \,.
\ee

When $\Delta\omega$ is not negligibly small, the probability for $M<M_0/2$ 
is not negligible, as it grows in proportion to $\dW$.
\citet{neistein06} showed that this effect is rather small 
and can be bounded by well defined limits on $q$, though the exact
value of $q$ is not fully specified by the EPS formalism.
They showed that \equ{app_acc} is still a valid approximation,
but with $q$ slightly larger than 2, and limited to the range 2-2.3 for
a flat $\Lambda$CDM cosmology with $0.1<\Omega_{\rm m}<0.9$ and
$10^8<M<10^{15}\msun$.  We adopt $q=2.2$ for all practical purposes.
 
As a sanity check, note that the obtained specific accretion rate at a
fixed mass is self-similar in time, consistent with the fact that the
only relevant mass scale at any given time is the non-linear clustering
mass of the Press-Schechter formalism, $\mps$, defined by
$\sigma[\mps(t)] = \delta_c/D(t)$. This is because the time dependence
in \equ{app_acc} is via $\dot\omega = - \sigma[\mps(t)]\dot{D}/D$.
Once $\dot{M}/M$ is divided by the universal growth rate $\dot{D}/D$,
the time dependence enters only through $\mps(t)$.
Any viable approximation for the cosmological accretion rate must obey 
a self-similarity of this sort.

We also note in \equ{sm} that at a fixed $t$, $\dot{M}/M \rightarrow \infty$
when $M \rightarrow \infty$, but recall that haloes of $M \gg \mps$ are rare,
such that most of the accreted mass is not onto arbitrarily massive haloes.

\subsection{Mass Growth History of $M_0$ at $t_0$}

The mass growth history of the main progenitor of a halo of mass $M_0$
at $t_0$ is obtained by integrating \equ{m_asm} over $\dd\omega$ and over
$\dd M$ in the two sides of the equation:
\be
\omega\!-\!\omega_0= F(M)\!-\!F(M_0) ,
\quad F(M) \equiv \int_M^\infty\!\! \frac{\dd m}{m\,s(m)} .
\label{eq:fm}
\ee
The desired average mass of the main progenitor at $t$
can be extracted from the above equation,
\be
M(t|M_0) = F^{-1}[\omega\!-\!\omega_0\, +F(M_0)] , \quad
w=\frac{\delta_c}{D(t)} ,
\label{eq:app_mph}
\ee
where $F^{-1}$ is the inverse function of $F$, to be evaluated
either explicitly or via numerical interpolation.

\subsection{A Power-Law Power Spectrum}

A simple, fully self-similar example is provided by a power-law power spectrum,
$P_k\propto k^n$, for which
\be
\sigma=\sigma_R M^{-\alpha} \,,
\quad \alpha\equiv (n+3)/6 \,.
\ee
Here $\sigma_R$ is a normalization constant,
and $M$ is measured in units of $M_R$, the average mass originally contained in
a top-hat sphere of comoving radius $R$ (e.g. $\sigma_8$ for $R=8\hmpc$).
In this case, the average accretion rate is given by \equ{app_acc} with
\be
s(M)=\tau M^\alpha \,,
\quad \tau \equiv \frac{1}{\sigma_R}
\left( \frac{2/\pi}{q^{2\alpha}-1} \right)^{1/2} \,.
\label{eq:sm1}
\ee
Then,
\be
F(M) = \frac{1}{\alpha\tau} (M^{-\alpha}-M_0^{-\alpha}) \,,
\ee
and the explicit solution for the average main-progenitor mass is
\be
M(t|M_0) = [M_0^{-\alpha} +\alpha\tau (\omega-\omega_0)]^{-1/\alpha}
\ee
(with $\omega = \delta_c/D(t)$, $M$ in units of $M_R$, and $q\simeq 2.2$).

\Equ{app_acc} with $s(M)$ from \equ{sm1} approximates
the accretion rate for a general power spectrum
once $\alpha$ is the local slope $\alpha(M) \equiv -d \ln \sigma / d \ln M$.
For $n=-2.1$, appropriate for $M\!\sim\! 10^{12}\msun$ in the $\Lambda$CDM
cosmology, we have $\alpha \simeq 0.15$. This is indeed similar to the
power index of
the mass dependence from N-body simulations \citep[][Fig.~12]{wechsler02}.

The time dependence can be approximated in standard $\Lambda$CDM by \equ{ddot}.
Using a crude power-law approximation for the growth of the Press-Schechter
mass out to $z\!\sim\! 2$, $M_*(z)/M_{*0} \simeq (1+z)^{-5}$, we end up with
a practical approximation for the accretion rate into haloes
in the vicinity of $M\!\sim\! 10^{12}\msun$,
\be
\frac{\dot M}{M} \simeq 0.04\, Gyr^{-1} (1+z)^{2.25} M_{12}^{0.15} \,,
\label{eq:app_acc_approx}
\ee
which is accurate to better than 10\% over the range $0\leq z\leq 2$.

\subsection{$\Lambda$CDM Power Spectrum}

For the standard $\Lambda$CDM power spectrum we adopt $P(k)=k\,T^2(k)$
with the transfer function \citep{bbks86}
\bea
T(k)\!\!\!\!\!&=&\!\!\!\!\!\frac{\ln(1+2.34\,\kappa)}{2.34\,\kappa} \\ 
\nonumber
\!\!\!\!\!&\times&\!\!\!\!\!\!\!\left[1\!+\! 3.89\,\kappa\!+\!(16.1\,\kappa)^2 
\!+\! (5.46\,\kappa)^3 \!+\! (6.71\,\kappa)^4
\right] ^{-1/4} . 
\eea
Here $\kappa \equiv k/\Gamma$,
with the wave number $k$ in units of $(\!\hmpc)^{-1}$,
and with the shape parameter $\Gamma$ \citep{sugiyama95}
\begin{equation}
\Gamma = \Omega_{\rm m} h\exp\left[-\Omega_{\rm b}(1+\sqrt{2h}/\Omega_{\rm m})
\right] \,.
\end{equation}
This power spectrum can be used to derive $\sigma(M)$ and the corresponding
$s(M)$ with $q=2.2$ in \equ{sm}. 
This can then be substituted in \equ{app_acc} for an explicit estimate of
the average accretion rate onto haloes of mass $M$ at $t$.  
It can then be used in \equ{fm} for computing
$F(M)$, evaluating its inverse function numerically, and using it in \equ{mph}
to obtain the average history of the main progenitor, $M(t|M_0)$.

Alternatively, following \citet{bosch02}, 
the rms fluctuation corresponding the $\Lambda$CDM
power spectrum is approximated (to better 1\% in the range
$10^6 < M <10^{16}\msun$) by 
\be
\sigma(M)=\sigma_8\, \frac {\Psi[(M/M_{\rm c})^{1/3}]} {\Psi[32\,\Gamma]} \,,
\ee
\be
M_{\rm c} = 9.73\times 10^{11}\msun\,
      h_{0.7}^{-1}\, (\Omega_{\rm m}/0.3)\, (\Gamma/0.2)^{-3} \,,
\ee
and with the fitting function
\bea
\Psi(x)\!\!\!\!\!&=&\!\!\!\!\!64.09\ ( 1 + 1.074\,x^{0.3}  - 1.581\,x^{0.4} 
\\ \nonumber & &\!\!\!\!\!  
  + 0.954\,x^{0.5} - 0.185\,x^{0.6} )^{-10} \,.
\eea

Substituting this analytic approximation in \equ{fm}, numerically
inverting $F(M)$, and using it in \equ{mph}, \citet{neistein06}
derived the following approximation for the average main progenitor
history:
\be
\frac{M(t|M_0)}{M_{\rm c}} =
F_q^{-1} \left[
\frac{\Psi(32\Gamma)}{\sigma_8}(\omega\!-\!\omega_0) +
F_q\left( \frac{M_0}{M_c} \right) \right] ,
\ee
with $q=2.2$ and the fitting function
\bea
F_{2.2}(u)\!\!\!\!\!&=&\!\!\!\!\!-6.92\!\times\! 10^{-5} (\ln u)^4 
+ 5.0\!\times\! 10^{-3} (\ln u)^3 
\\ \nonumber & &\!\!\!\!\! 
+8.64\!\times\!10^{-2} (\ln u)^2 
- 12.66\, (\ln u) 
  \,.
\eea
This fitting function is accurate to better than 1\% over the range
$10^{6} < M < 10^{15}\msun$ for the standard $\Lambda$CDM cosmology.

\subsection{In Cosmological N-body Simulations}
\label{sec:wec}

\citet{wechsler02} analyzed an N-body simulation of $\Lambda$CDM for
the average mass-growth history of the 
main progenitor of a halo of mass $M_0$ at $t_0$. They found a good fit with
\be
M(t|M_0) = M_0\, e^{-2 a_{\rm c} z} \,,
\label{eq:wec}
\ee
\be
a_{\rm c} = 0.28\, \left(\frac{M_0}{10^{12}\msun}\right)^{0.15} \,. 
\ee
This function is a good practical approximation to the EPS estimate
specified above, as long as $t_0$ is the present time, $z=0$.

\subsection{Useful Relations for Linear Fluctuations}

Following\citet{mo02}, in a flat universe
\be
D(z)=\frac{g(z)}{g(0)} \frac{1}{(1+z)} \,, 
\ee
\be
g(z) \simeq \frac{5}{2} \left( \Omega_m^{-3/7}(z)+\frac{3}{2}\right)^{-1} \,,
\ee
with a negligible error on the order of $\Omega_\lambda/70$ in $g(z)$.
Thus
\be
\frac{\dot D}{D} \simeq H(z) \left(
1+\frac{18}{35}\, g(z)\, \Omega_m^{-3/7}(z)\, [\Omega_m(z)-1] \right) \,.
\ee
Here
\be
H(z)=H_0 E(z), \quad \Omega_m(z) = \frac{\Omega_{m0}(1+z)^3}{E^2(z)} \,,
\ee
\be
E^2(z)=1-\Omega_{m0}+\Omega_{m0}(1+z)^3 \,.
\ee

For $\Lambda$CDM with $\Omega_{\rm m}=0.3$ and $\Omega_{\Lambda}=0.7$,
a practical approximation is
\be
\frac{\dot D}{D} \simeq 0.0385\, (1+z)^{3/2}\, h_{0.7}\,Gyr^{-1} \,,
\label{eq:ddot}
\ee
which is accurate to better than 1\% at $z\geq 1$ and better than
10\% at $z \geq 0$.

The fluctuation variance is derived in the general case by integrating
the power spectrum,
\be
\sigma^2(M)=\frac{1}{2\pi} \int_0^\infty dk\, k^2\, P(k)\, {\tilde{W}}^2(kR)\,,
\ee
with $M=(4\pi/3) \rho_0 R^3$,
and with the Fourier transform of the real-space top-hat window function
or radius $R$
\be
\tilde{W}(x)=3(\sin x -x\cos x)/x^3 \,.
\ee

\section{Recipes for SAMs}
\label{sec:app_sam}

\subsection{Following Halo Evolution in SAM}
\label{sec:app_sam_sam}

Given the mass growth of each halo in the merger tree, $\Mv(t)$,
the SAMBA onset time $t_1$ (redshift $z_1$)
is identified by $\Mv(t_1)=M_1=10^{12}\msun$.
The SAMBA is classified as ``early type" or ``late type" according to
whether $z_1>2.5$ or $z_1\leq 2.5$ respectively.
Then the beginning and the end of the burst, as read from \fig{sumtl}, are
\be
t_2=
\cases{&$\!\!\!\!\! 1.80\, t_1 ,\quad z_1>2.5$\cr
       &$\!\!\!\!\! 1.36\, t_1 ,\quad z_1\leq 2.5$ } \ ,
\quad t_3=1.18\, t_2 \ .
\ee
This defines the four phases of gas cooling.

The growth rates of cold and hot gas mass during each phase
can be approximated as follows:

\smallskip\no
1. {\bf Cold-flow} phase, $t \leq t_1$,
\be
\Msd=\fb \Mvd \ ,
\quad \Mhd=0 \ .
\ee

\smallskip\no
2. {\bf Tentative-quenching} phase, $t_1<t \leq t_2$,
\be
\Msd = \fb\, \frac{\Delta {\Mv}_{12}}{\Delta t_{12}} \times 
\cases{&$\!\!\!\!\! 0.08, \quad z_1>2.5$\cr
       &$\!\!\!\!\! 0.3, \quad\ \,  z_1\leq 2.5$ }  ,
\ee
\be
\Mhd = \fb \Mvd - \Msd \ ,
\ee
where $\Delta {\Mv}_{ij} = \Mv (t_j)-\Mv (t_i)$ and $\Delta t_{ij}=t_j-t_i$.
The numerical factors for early and late SAMBAs, both for the tentative 
quenching and for the burst below,
are read from straightforward fits to $\Delta \Ms/(\Delta \Mv \fb)$ in
\fig{sumv}.

\smallskip\no
3. {\bf Burst} phase, $t_2<t \leq t_3$,
\be
\Msd = \fb\, \frac{\Delta {\Mv}_{23}}{\Delta t_{23}} \times
\cases{&$\!\!\!\!\! 0.5, \quad z_1>2.5$\cr
       &$\!\!\!\!\! 1.6, \quad z_1\leq 2.5$ }  \,,
\ee
\be
\Mhd = - \frac{{\Mh}_{2}}{\Delta t_{23}} \,,
\ee
where ${\Mh}_2$ is the accumulated hot gas mass by $t_2$,
${\Mh}_2 = \fb \Delta{\Mv}_{12} -\dot{M}_{\rm disc,2} \Delta t_{12}$.
The cold accretion is more efficient in the late bursts because of the
tentative shock instability due to the lower virial accretion rate at
late times.

\smallskip\no
4. {\bf Post-burst} phase, $t>t_3$,
\be
\Msd = \fb \Mvd \times
\cases{&$\!\!\!\!\! 0.12,\quad z_1>2.5$\cr
       &$\!\!\!\!\! 0, \quad\quad\ z_1\leq 2.5$ } \,,
\ee
\be
\Mhd = \fb \Mvd - \Msd  .
\ee
The numerical factor for early SAMBAs is read from the fit
$\Mv/\Ms \simeq 4/\fb$ at to the results shown in
\fig{sumv}, combined with the fact, seen in \fig{sf57}, that during the
post-burst phase $\Msd/\Ms \simeq \Mvd/\Mv$ and $\Mv/\Ms \simeq const$.
Recall that the early SAMBAs, ending up in $\Mv \geq 10^{13}\msun$ haloes
today,
should be quenched after $t_3$ by another mechanism, such as clumpy
accretion or AGN feedback. If such a process is not explicitly
included in the SAM,
one can simply set $\Msd=0$ in the post-burst phase for all haloes \citep[as
in][]{cattaneo06}.

The disc mass growth according to the proposed SAMBA recipe for SAM is
shown in \fig{sam}, in comparison with the
standard disc growth following the evolution of a ``cooling radius".

\subsection{SFR as a function of Halo Mass: FHM}
\label{sec:app_sam_fhm}

An alternative, quick and crude way to learn about the implications of
SAMBAs is by providing a recipe for the SFR as a function of halo mass
and redshift, to be incorporated in a given halo population in a SAM while
bypassing the detailed baryonic processes of cooling, star formation,
and feedback \citep[Formation History Modeling, FHM,][]{croton07}.
The fact that the SAMBA phenomenon is driven by halo mass allows us to
provide a concrete recipe of this sort.

\begin{figure}
\vskip 7.1cm
\includegraphics{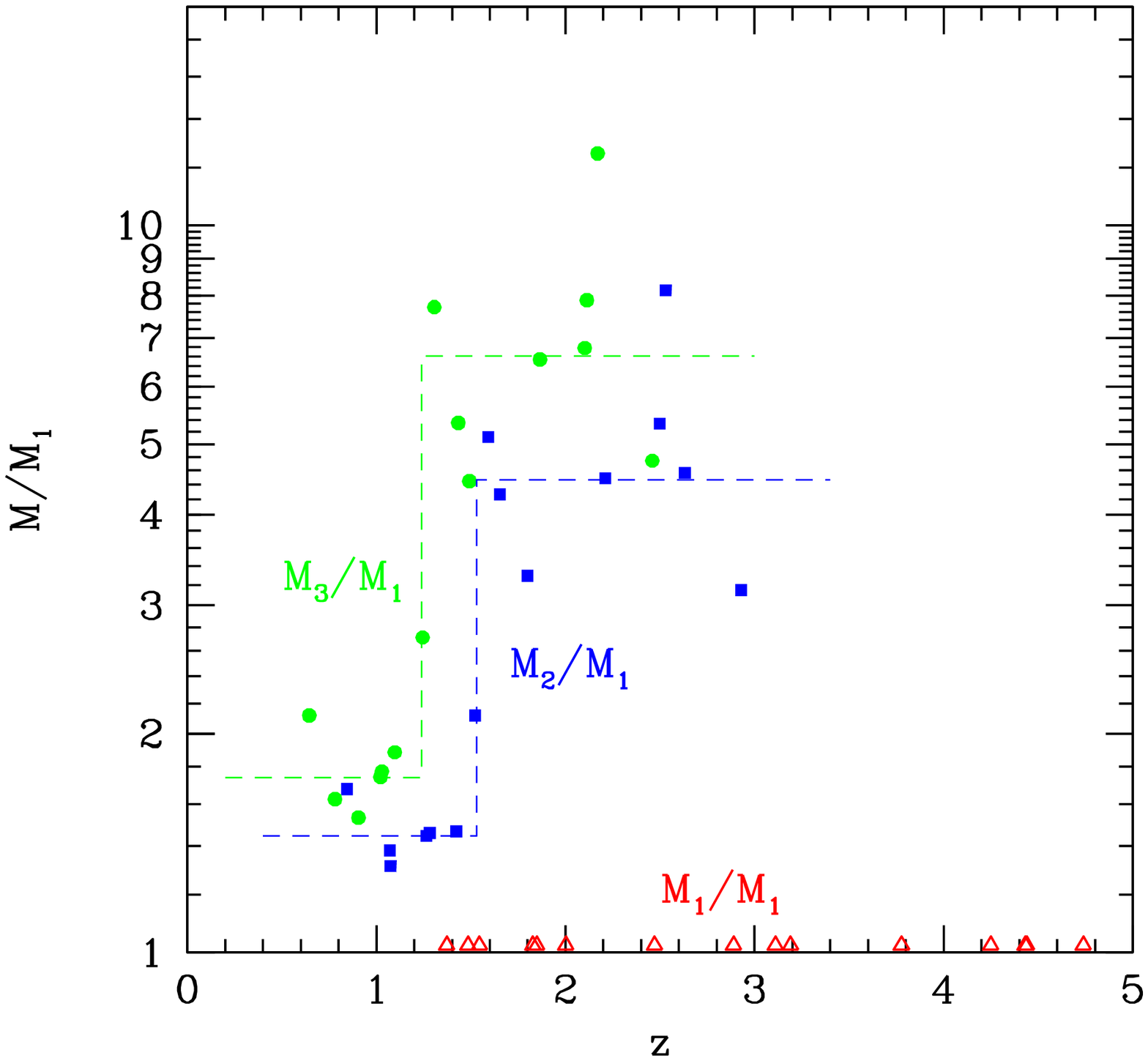}
\caption{The characteristic SAMBA masses relative to $M_1\simeq 10^{12}\msun$.
Open red triangles refer to $z_1$.
Solid blue squares refer to $M_2/M_1$ at $z_2$.
Solid green circles refer to $M_3/M_1$ at $z_3$.
The lines mark the fits used for the FHM recipe.
}
\label{fig:mv}
\end{figure}

The SAMBA characteristic halo masses $M_1$, $M_2$ and $M_3$,
corresponding to the characteristic times $t_1$, $t_2$ and $t_3$,
can be approximated based on our simulations by
\be
M_1 =10^{12}\msun \,,
\ee
\be
M_2/M_1 =
\cases{&$\!\!\!\!\! 4.45  ,\quad z>1.53$\cr   
       &$\!\!\!\!\! 1.45  ,\quad z\leq 1.53$} \,,
\ee
\be
M_3/M_1 =
\cases{&$\!\!\!\!\! 6.61  ,\quad z>1.24$\cr
       &$\!\!\!\!\! 1.74  ,\quad z\leq 1.24$ }\,.
\ee
The constant value of $M_1$ is as predicted by \citet{db06} for
a realistic gradual increase of metallicity with time.
Then $M_2$ and $M_3$ are obtained from the fits shown in \fig{mv}.
Notice the interesting prediction that in the range $1.24 < z \leq 1.53$
$M_3$ is much larger than $M_2$, giving rise to a broad range of massive
haloes with high SFR.

Given the virial accretion rate in the provided merger tree, $\Mvd$,
the disc accretion rate, which could be interpreted as an upper limit for
the SFR, is approximated in the different mass ranges by:

\smallskip\no
1. {\bf Cold-flow} range, $\Mv \leq M_1$, 
\be
\Msd = \fb\,\Mvd \ .
\ee

\smallskip\no
2. {\bf Tentative-quenching} range, $M_1 < \Mv \leq M_2$,
\be
\Msd= \fb\,\Mvd \times
   \cases{&$\!\!\!\!\! 0.08  ,\quad z >1.8$\cr
          &$\!\!\!\!\! 0.3  ,\quad\ \, z\leq 1.8$ } \ .
\ee

\smallskip\no
3. {\bf Burst} range, $M_2 < \Mv \leq M_3$,
\be
\Msd= \fb\,\Mvd \times
   \cases{&$\!\!\!\!\! 0.5    ,\quad z >1.4 $\cr
          &$\!\!\!\!\! 1.6    ,\quad z\leq 1.4$ } \ .
\ee
The numerical values are read
from crude fits to the simulation results for $\Delta \Ms/(\Delta \Mv \fb)$,
as shown in \fig{sumv}. 

\smallskip\no
4. {\bf Post-burst} range, $\Mv > M_3$, 
\be
\Msd= \fb\,\Mvd \times
   \cases{&$\!\!\!\!\! 0.12    ,\quad \Mv > M_{\rm q}(z)$\cr
          &$\!\!\!\!\! 0       ,\quad\quad \ \Mv \leq M_{\rm q}(z)$ } \ ,
\ee
\be
M_{\rm q}(z) \simeq 10^{13}\msun \, \exp(-0.75\,z) \,.
\ee
The upper-limit for effective quenching, $M_{\rm q}(z)$, is the 
mass of the intermediate SAMBA separating late from early SAMBAs, 
which eventually ends up as $\Mv = 10^{13}\msun$ today.
The above estimate is based on the fit to simulations, \equ{wec}.  
Note that once $z>1.4$, $M_3 > M_{\rm q}(z)$, 
so in this redshift range $\Mv > M_3$ automatically implies $\Mv > M_{\rm q}$.

The numerical values in the quenching and burst phases are read
from the crude fits to the simulation results shown in \fig{sumv},
in most of which the virial mass grows according to the average 
cosmological rate. When implementing the above recipe, one can either use
the actual mass growth rate of the haloes drawn from merger-tree realizations,
or the average virial accretion rate as provided in \equ{acc_approx}. 

The main feature of the model is the quenching of SFR above a threshold
halo mass, $\Msh\! \sim\! 10^{12}\msun$, as already implemented in SAM
simulations \citep{cattaneo06,croton06}. 
The interesting new feature predicted by the model is the high SFR in
a second range of masses above $\Msh$. At low redshifts, this is a 
slight effective increase of the threshold mass to 
$\sim\! 2\times 10^{12}\msun$.
At $z\geq 1.3$, the high SFR is predicted to extend to haloes as large as
$\sim\! 7\times 10^{12}\msun$. This should allow the appearance of massive blue
galaxies at high redshifts.
In the special range $1.24< z <1.53$, the high SFR is expected over an
especially broad range of masses. Given that $10^{12}\msun$ is the typical
mass for haloes that form at this redshift range, this should lead to
a peak in the history of cosmological SFR density (the ``Madau" plot).

\subsection{Fine-tuning of the SAMBA Recipes}

While the above recipes are our best guess based on the spherical
model, with no free parameters, a better match to observations may 
require modest fine-tuning of model parameters, e.g., allowing for 
certain uncertainties in the assumptions made in the computations, 
as well as compensating for deviations from spherical symmetry and 
uniform accretion. 

The uncertainty in the value of the critical mass for shock heating, $\Msh$,
can easily accommodate a factor of two and some redshift dependence.
As computed in \citet{db06}, it is a function
of certain physical parameters, in particular redshift $z$ and the average 
metallicity $Z$ and gas fraction $\fb$ in the inner halo, which also vary with
$z$. The estimates of $\Msh$ as a function of $z$ are shown in their figure~2.
A practical approximation based on their equation~34 is
\be
\Msh \simeq 10^{12}\msun\,
\frac
{Z_{0.2}^{0.52} {\fb}_{0.05}^{3/4}}
{(1+z)^{3/8}}
\left(\frac{\ln(1+C_0)}{\ln(1+C)}\right)^{3/4} ,
\ee
where $Z_{0.2}\equiv Z/(0.2Z_\odot)$ and ${\fb}_{0.05} \equiv \fb/0.05$.
In this formula, following \citet{db06}, the variation of mean metallicity 
with redshift can be crudely approximated by
\be
\log \frac{Z}{Z_0} = -0.17\, z \,.
\ee
Following \citet{bullock01_c}, the average halo concentration near 
$\Mv\!\sim\! 10^{12}\msun$ is evolving as 
\be
C=C_0\,(1+z)^{-1} \,, \quad C_0 \simeq 13 \,.
\ee
With this choice, the critical mass is $\Msh\!\sim\!10^{12}\msun$ and is only
a weak function of redshift.
It becomes even weaker if $\fb$ gets slightly higher at high $z$. 
One can see that a factor of 4 change in the unknown $Z_0$ results in a 
factor of $\sim\!2$ in $\Msh$.

Another uncertainty, which is explicit in the FHM recipe, 
is in the actual SFR that is 
associated with the maximum possible rate provided by the disc accretion rate. 
This can be modeled as a free ratio, smaller than unity, between the two.

\label{lastpage}
\end{document}